\documentclass[12p]{article}
\usepackage{latexsym}
\usepackage{amssymb}
\usepackage{color}
\usepackage{amsmath}
\usepackage{epsfig}
\usepackage{hyperref}
 \usepackage{dcolumn}
   \usepackage{threeparttable}
     \newcommand{\thickhline}{\noalign{\hrule height 0.8pt}}
\newcommand{\ortala}[1]{\begin{center}#1\end{center}}

\newcommand{\sandd}[1]{\left\langle #1\right\rangle}

\newcommand{\integ}[3]{{{\underset{#1 }{\overset{#2}{\displaystyle\int}}}#3}}

\newcommand{\summ}[3]{{{\underset{#1 }{\overset{#2}{\displaystyle\sum}}}#3}}
\newcommand{\prodd}[3]{{{\underset{#1
}{\overset{#2}{\displaystyle\prod}}}#3}}

\newcommand{\re}[1]{(\ref{#1})}

\newcommand{\eq}[2]{\begin{equation}\label{#1}  #2\end{equation}}

\newcommand{\paran}[1]{\left(#1\right)}

\newcommand{\sch}[1]{Schrodinger}

\newcommand{\komb}[2]{\paran{\begin{array}{c} #1 \\ #2 \end{array}}}

\linethickness{0.6pt}

\begin{document}

\ortala{\textbf{Bond diluted anisotropic quantum Heisenberg model}}

\ortala{\textbf{\"Umit Ak\i nc\i \footnote{umit.akinci@deu.edu.tr}}}

\ortala{\textit{Department of Physics, Dokuz Eyl\"ul University,
TR-35160 Izmir, Turkey}}

\section{Abstract}
Effects of the bond dilution on the critical temperatures,  phase
diagrams and the magnetization behaviors of the isotropic and anisotropic quantum Heisenberg model have been
investigated in detail. For the isotropic case, bond percolation threshold values have been determined for several numbers of two (2D) and three (3D) dimensional lattices. In order to investigate the effect of the anisotropy in the exchange interaction on the results obtained for the isotropic model, a detailed investigation has been made on a honeycomb lattice. Some interesting results, such as second order reentrant phenomena in the phase diagrams have been found.
Keywords: \textbf{Quantum anisotropic Heisenberg model; bond dilution; bond percolation threshold}

\section{Introduction}\label{introduction}
Quenched randomness effects are very important in modeling real materials, since consideration of these effects simulates a realistic model of real materials. Real materials have some uncontrollable defects and these defects can be modeled by introducing site dilution (randomly distributed non magnetic atoms), bond dilution (randomly broken bonds between the magnetic atoms) or both of them into the related model. It is a well known fact that, Heisenberg model produces more realistic results than the Ising model, in order to explain the magnetic properties of real materials. Thus, it is important to work on the Heisenberg model with these quenched randomness effects.
These quenched randomness effects produce different behaviors in the magnetic properties of the model, e.g. different phase transition characteristics from the pure model (i.e. the model without any quenched randomness effects).

Heisenberg model with quenched randomness effects has been studied widely by a variety of methods, such as spin-1/2 (S-1/2) isotropic  Heisenberg model with bond dilution on 2D lattices with Monte Carlo (MC) simulation  \cite{ref1,ref2,ref3,ref4}, anisotropic model (by means of the XXZ model) on 2D lattices with real space renormalization group (RSRG) technique \cite{ref5}. On the other hand, random bond distributed systems in which spin glass phases originate have been studied, e.g. discrete distribution on S-1/2 Heisenberg model with pair approximation \cite{ref6,ref7}, density matrix product approximation \cite{ref8}, Gaussian distribution with imaginary time Grassmann field theory \cite{ref9} and S-1 Heisenberg model with discretely distributed random bonds with exact diagonalization method\cite{ref10}. Besides, both site and bond diluted systems have been studied with quantum MC on 2D lattices \cite{ref11} and S-1/2 Heisenberg model with site-bond correlated dilution (which covers uncorrelated site dilution as a limit) on 2D and 3D lattices with RSRG \cite{ref12} and MC \cite{ref13}, and also using a variational principle for the free energy  \cite{ref14}. All of these work are related to the isotropic Heisenberg model or XXZ model.

The aim of this work is determine the effect of the bond dilution on the phase diagrams and the thermodynamic properties of the anisotropic quantum Heisenberg model. By anisotropy, we do not restrict ourselves with XXZ model. Namely, we want to determine the effect of the anisotropy in the exchange interaction on the phase transition characteristics not only by means of XXZ model.  The method is effective field theory (EFT) with two spin cluster approximation\cite{ref15}. EFT approximation can provide results that are superior to those obtained within the traditional mean field approximation (MFA), due to the consideration of self spin correlations which are omitted in the MFA.

EFT for a typical magnetic system starts by constructing a finite cluster of spins which represents the system. Callen-Suzuki spin identities \cite{ref16,ref17} are the starting point of the EFT for the one spin clusters. If one expands these identities with differential operator technique\cite{ref17ek}, multi spin correlations appear, and in order to avoid from the mathematical difficulties, these multi spin correlations are often neglected by using decoupling approximation (DA) \cite{ref18}. Working with larger finite clusters will give more accurate results. Callen-Suzuki identities have been generalized to two spin clusters in Ref. \cite{ref19} (namely EFT-2 formulation). This EFT-2 formulation has been successfully applied to a variety of systems, such as quantum S-1/2 Heisenberg ferromagnet \cite{ref20,ref21}  and antiferromagnet \cite{ref22} systems,  classical n-vector model \cite{ref23,ref24}, and spin-1 Heisenberg ferromagnet \cite{ref25,ref26}.

The paper is organized as follows: In Sec. \ref{formulation}, we
briefly present the model and  formulation. The results and
discussions are presented in Sec. \ref{results}, and finally Sec.
\ref{conclusion} contains our conclusions.

\section{Model and Formulation}\label{formulation}

We consider a lattice which consists of $N$ identical spins (S-$1/2$) such that each of the spins has $z$ nearest neighbors. The Hamiltonian of the system is given by
\eq{denk1}{\mathcal{H}=-\summ{<i,j>}{}{\paran{J_x^{(ij)} s_i^xs_j^x+J_y^{(ij)} s_i^ys_j^y+J_z^{(ij)} s_i^zs_j^z}}}
where $s_i^x,s_i^y$ and  $s_i^z$ denote the Pauli spin operators at a site $i$. $J_x^{(ij)},J_y^{(ij)}$ and $J_z^{(ij)}$ stand for the components of the exchange interaction $\mathbf{J}$ (in other words anisotropy in the exchange interaction) between the nearest neighbor spins $i$ and $j$ . The sum is carried over the nearest neighbors of the lattice. The bonds between the spins $i$ and $j$ are randomly distributed via
\eq{denk2}{P\paran{\mathbf{J}^{(ij)}}=c\delta\paran{\mathbf{J}^{(ij)}-\mathbf{J}}+\paran{1-c}\delta\paran{\mathbf{J}^{(ij)}}
} in the bond dilution problem. The distribution given by Eq. \re{denk2} distribute bonds randomly between lattice sites as $c$ percentage of bonds are closed and remaining $1-c$ percentage of bonds are open, i.e. $c$ is the concentration of closed bonds in the
lattice. Here $\delta$ stands for the delta function and $c$ is a real number which is defined within the range of $0\le c \le 1$.
The distribution given by Eq. \re{denk2}
reduces to the system with homogenously distributed bonds (i.e. pure system) for $c=1$.

We use the two spin cluster approximation as an EFT formulation, namely EFT-2 formulation\cite{ref15}. In this approximation, we choose two spins (namely $s_1$ and $s_2$) and treat the interactions in this two spin cluster exactly. In order to avoid some mathematical difficulties we replace the perimeter spins of the two spin cluster by Ising spins (axial approximation) \cite{ref24}. With the procedure defined in Ref.
\cite{ref24}, we get an expression for the magnetization per spin as
\eq{denk3}{
m=\sandd{\sandd{\frac{1}{2}\paran{s_1^z+s_2^z}}}_r=\sandd{\sandd{\frac{h_{+}^\prime}{X_0^\prime}\frac{\sinh{\paran{\beta X_0^\prime}}}{\cosh{\paran{\beta X_0^\prime}}+\exp{\paran{-2\beta J^{(12)}_z}}\cosh{\paran{\beta Y_0^\prime}}}}}_r
} where $\beta=1/(k_B T)$, $k_B$ is Boltzmann
constant and $T$ is the temperature. The inner average bracket in Eq. \re{denk3} (which has no
subscript) stands for thermal average and the outer one (which has subscript r) is for the configurational
averaging which is necessary for including the effect of the random bond distribution. The parameters in Eq. \re{denk3} are given by
\eq{denk4}{
\begin{array}{lcl}
X_0^\prime&=&\left[\paran{J^{(12)}_x-J^{(12)}_y}^2+h_{+}^{\prime 2}\right]^{1/2}\\
Y_0^\prime&=&\left[\paran{J^{(12)}_x+J^{(12)}_y}^2+h_{-}^{\prime 2}\right]^{1/2}\\
\end{array}
}
\eq{denk5}{
\begin{array}{lcl}
h_{+}^{\prime 2}&=&h_1^\prime+h_2^\prime\\
h_{-}^{\prime 2}&=&h_1^\prime-h_2^\prime\\
\end{array}
}
\eq{denk6}{
\begin{array}{lcl}
h_1^\prime&=&\summ{k}{}{J^{(1k)}_zs_{k}^z}\\
h_2^\prime&=&\summ{l}{}{J^{(2l)}_zs_{l}^z}
\end{array}
} where $s_k^z$ stands for the $z$ component of the nearest neighbor of the spin $s_1$ while $s_l^z$ stands for the $z$ component of the nearest neighbor of the spin $s_2$. The sums in the Eq. \re{denk6} are over the nearest neighbor sites of the sites labeled $1$ and $2$, respectively.
The configurational averages
can be calculated via integration of the expression  by using Eq. \re{denk2}, over the all bonds of the treated cluster.
\eq{denk7}{
m=\sandd{\integ{}{}{}d\mathbf{J}^{(12)}P\paran{\mathbf{J}^{(12)}}\prodd{k,l}{}{}d\mathbf{J}^{(1k)}
d\mathbf{J}^{(2l)}P\paran{\mathbf{J}^{(1k)}}P\paran{\mathbf{J}^{(2l)}}\frac{h_{+}^\prime}{X_0^\prime}\frac{\sinh{\paran{\beta X_0^\prime}}}{\cosh{\paran{\beta X_0^\prime}}+\exp{\paran{-2\beta J^{(12)}_z}}\cosh{\paran{\beta Y_0^\prime}}}}
} where product is taken over the nearest neighbors of the sites $1$ and $2$.
If we perform the integration concerning the bond between the sites $1$ and $2$ then we get the following expression,
$$
m=\left\langle\integ{}{}{}\prodd{k,l}{}{}d\mathbf{J}^{(1k)}
d\mathbf{J}^{(2l)}P\paran{\mathbf{J}^{(1k)}}P\paran{\mathbf{J}^{(2l)}}
h_{+}^\prime\times \right.
$$
\eq{denk8}{
\left.
\left[\frac{c}{X_0^{\prime\prime}}\frac{\sinh{\paran{\beta X_0^{\prime\prime}}}}{\cosh{\paran{\beta X_0^{\prime\prime}}}+\exp{\paran{-2\beta J_z}}\cosh{\paran{\beta Y_0^{\prime\prime}}}}+
\frac{1-c}{h_{+}^\prime}\frac{\sinh{\paran{\beta h_{+}^\prime}}}{\cosh{\paran{\beta h_{+}^\prime}}+\cosh{\paran{\beta h_{-}^\prime}}}
\right]\right\rangle
}
where
\eq{denk9}{
\begin{array}{lcl}
X_0^{\prime\prime}&=&\left[\paran{J_x-J_y}^2+h_{+}^{\prime 2}\right]^{1/2}\\
Y_0^{\prime\prime}&=&\left[\paran{J_x+J_y}^2+h_{-}^{\prime 2}\right]^{1/2}.\\
\end{array}
}
Eq. \re{denk8} can be written in terms of the differential operators using differential operator technique \cite{ref17ek} and it is found as
\eq{denk10}{
m=\left\langle\integ{}{}{}\prodd{k,l}{}{}d\mathbf{J}^{(1k)}
d\mathbf{J}^{(2l)}P\paran{\mathbf{J}^{(1k)}}P\paran{\mathbf{J}^{(2l)}}\exp{\paran{h_{1}^\prime\nabla_x}}\exp{\paran{h_{2}^\prime\nabla_y}}\right\rangle f\paran{x,y}|_{x=0,y=0}
}
where
\eq{denk11}{
f\paran{x,y}=\frac{cz_{+}}{X_0}\frac{\sinh{\paran{\beta X_0}}}{\cosh{\paran{\beta X_0}}+\exp{\paran{-2\beta J_z}}\cosh{\paran{\beta Y_0}}}+
\paran{1-c}\frac{\sinh{\paran{\beta z_{+}}}}{\cosh{\paran{\beta z_{+}}}+\cosh{\paran{\beta z_{-}}}}
}
and

$$
X_0=\left[\paran{J_x-J_y}^2+z_{+}^{ 2}\right]^{1/2}
$$
\eq{denk12}{
Y_0=\left[\paran{J_x+J_y}^2+z_{-}^{ 2}\right]^{1/2}
}
$$
z_{+}=x+y,\quad z_{-}=x-y.
$$

In Eq. \re{denk10}, the parameters $\nabla_x=\partial/\partial x$ and $\nabla_y=\partial/\partial y$ are the usual differential operators in the
differential operator technique. Differential operators act on an arbitrary function via
\eq{denk13}{\exp{\paran{a\nabla_x+b\nabla_y}}g\paran{x,y}=g\paran{x+a,y+b}}
with any constant  $a$ and $b$.

Now let us assume that each of $s_1$ and $s_2$ has number of $z_0$ distinct nearest neighbors and both of them have $z_1$ common nearest neighbors. This means that, in Eq. \re{denk6}, the two sums have number of $z_1$ common terms.  If we take the integration in Eq. \re{denk10} with the help of Eqs. \re{denk2} and \re{denk6}  and using DA \cite{ref18}
we get an expression,
\eq{denk14}{
m=\sandd{\left[A_{x}+m B_{x}\right]^{z_0}
\left[A_{y}+m B_{y}\right]^{z_0}
\left[A_{xy}+m B_{xy}\right]^{z_1}} f\paran{x,y}|_{x=0,y=0}
} for the magnetization. The coefficients are defined by
\eq{denk15}{
\begin{array}{lcl}
A_{x}=c\cosh{\paran{J_z\nabla_x}}+(1-c)&\quad&
B_{x}=c\sinh{\paran{J_z\nabla_x}}\\
A_{y}=c\cosh{\paran{J_z\nabla_y}}+(1-c)
&\quad&
B_{y}=c\sinh{\paran{J_z\nabla_y}}\\
A_{xy}=c\cosh{\left[J_z\paran{\nabla_x+\nabla_y}\right]}+(1-c)&\quad&
B_{xy}=c\sinh{\left[J_z\paran{\nabla_x+\nabla_y}\right]}.\\
\end{array}
}

With the help of the Binomial expansion, Eq. \re{denk14} can be written as
\eq{denk16}{
m=\summ{p=0}{z_0}{}\summ{q=0}{z_0}{}\summ{r=0}{z_1}{}C^\prime_{pqr}m^{p+q+r}
} where the coefficients are
\eq{denk17}{
C^\prime_{pqr}=\komb{z_0}{p}\komb{z_0}{q}\komb{z_1}{r}A_x^{z_0-p}A_y^{z_0-q}A_{xy}^{z_1-r}B_x^{p}B_y^{q}B_{xy}^{r}f\paran{x,y}|_{x=0,y=0}
} and these coefficients can be calculated by using the definitions given in Eqs. \re{denk11},\re{denk13} and \re{denk15}. Let us write Eq. \re{denk16} in more familiar form as
\eq{denk18}{
m=\summ{k=0}{z}{}C_{k}m^{k}
}
\eq{denk19}{
C_{k}=\summ{p=0}{z_0}{}\summ{q=0}{z_0}{}\summ{r=0}{z_1}{}\delta_{p+q+r,k}C^\prime_{pqr}
} where $\delta_{i,j}$ is the Kronecker delta.

For a given set of Hamiltonian parameters ($J_x,J_y,J_z$), temperature and bond distribution parameter ($c$), we can determine the coefficients  from Eq. \re{denk17} and we can obtain
a non linear equation from Eq. \re{denk18}. By solving this equation, we can get the magnetization ($m$) for a given set of parameters and temperature. Since the magnetization is close to zero in the vicinity of the critical point then we can obtain a linear equation by linearizing the equation given in  Eq. \re{denk18} which allows us to determine the critical temperature. Since we have not calculated the free energy in this approximation, we can locate only second order transitions from the condition given as
\eq{denk20}{
C_1=1, \quad C_3<0.
}The tricritical point at which second and first order transition lines meet can be determined from the condition
\eq{denk21}{
C_1=1, \quad C_3=0.
}

\section{Results and Discussion}\label{results}
Let us scale the exchange interaction components with the  unit of energy  $J$ as
\eq{denk22}{
r_n=\frac{J_n}{J}
 } where $n=x,y,z$. Let us choose $r_z=1$, then $r_x,r_y$ can be used as the measure of the anisotropy in the exchange interaction. It can be seen from the formulation of the problem (e.g. see Eqs. \re{denk11}-\re{denk15}) that the transformation $J_x\rightarrow J_y,J_y\rightarrow J_x$ does not change the expression of the order parameter, regardless of the bond distribution. Hence, we can say that aforementioned transformation does not affect the formulation. Therefore, let us fix $r_x$  and concentrate only on the varying $r_y$ values. Our investigation will be focused on selected several 2D and 3D lattices in isotropic case and on a honeycomb lattice in anisotropic case with the ferromagnetic exchange interaction, i.e. $r_x,r_y,r_z>0$. It can be seen from the Hamiltonian of the system which is given by Eq. \re{denk1} that for $r_z > 0$, $r_x=r_y=0$ the system reduces to a model which has Ising type symmetry and $r_x=0,r_y,r_z > 0$ or $r_x,r_z > 0,r_y=0$ reduces the system to the system that have $XY$ type symmetry.

\subsection{Isotropic Model}

In this case, all components of the exchange interaction are equal to each other, i.e. $r_n=1$ in Eq. \re{denk22}. In order to represent the  behavior of the critical temperature ($k_BT_c/J$) as a function of the bond concentration ($c$) for the isotropic quantum Heisenberg model, we depict the phase diagrams for three different lattices in ($k_BT_c/J,c$) plane in Fig. \re{sek1}(a). As we can see from Fig. \re{sek1}(a) that the critical temperature of the system gradually decreases when the bond concentration decreases and takes the value of zero at a specific value of bond concentration, namely at the bond percolation threshold ($c^*$) which depends on the lattice geometry. At this point, we have to mention that bond percolation threshold value may not correspond to the value that makes the critical temperature zero. If the phase diagram of the system exhibits a reentrant behavior in the ($k_BT_c/J,c$) plane then the bond percolation threshold value corresponds to the $c$ value that terminates the phase diagram, i.e. for the value that provide $c<c^*$ system can not reach the ordered phase any more. In the isotropic case of the quantum Heisenberg model, the whole phase diagrams in the ($k_BT_c/J,c$) plane do not exhibit reentrant behavior.

The numerical values of bond percolation threshold values obtained within the present work are  $c^*=0.5903$ for a honeycomb lattice,  $c^*=0.4427$ for a square lattice, and $c^*=0.2974$ for a simple cubic lattice. We can also see from  Fig. \re{sek1}(a) that at a certain $c$, simple cubic lattice has highest critical temperature value which is due to the large number of  the nearest neighbor spins in the lattice. Besides, in Fig. \re{sek1}(b) we can see the variation of the ground state magnetization ($m_0$) with the bond concentration for three lattice geometries. These values of magnetizations calculated at the temperature $k_BT/J=0.001$ can be treated as ground states since the energy comes from the thermal fluctuations are negligible in comparison with the energy comes from the spin-spin interaction. As we can see from  Fig. \re{sek1}(b)  that, ground state magnetization starts from the value of $m_0=1.0$ and, stays in this value for a while when the bond concentration decreases. After then, it starts to fall and reach the zero at a bond percolation threshold value.	  	

\begin{figure}[h]\begin{center}
\epsfig{file=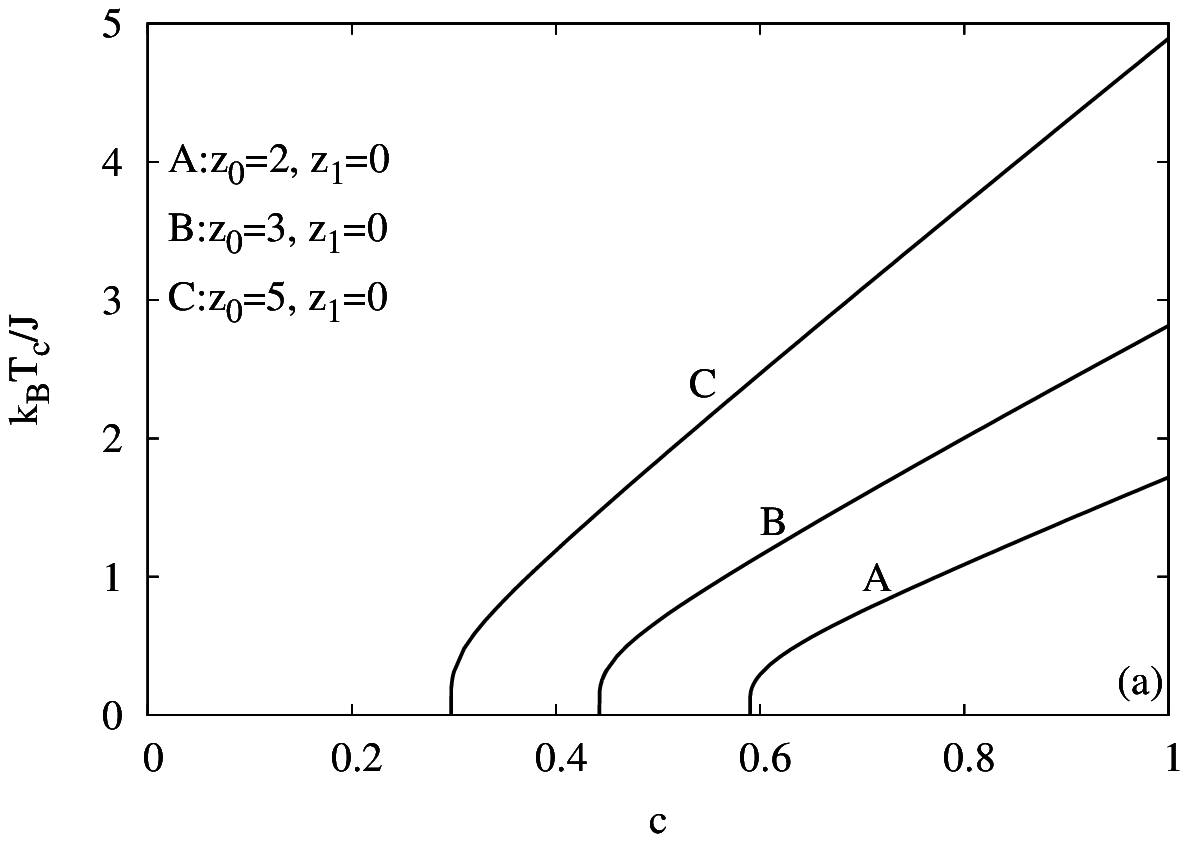, width=7.0cm}
\epsfig{file=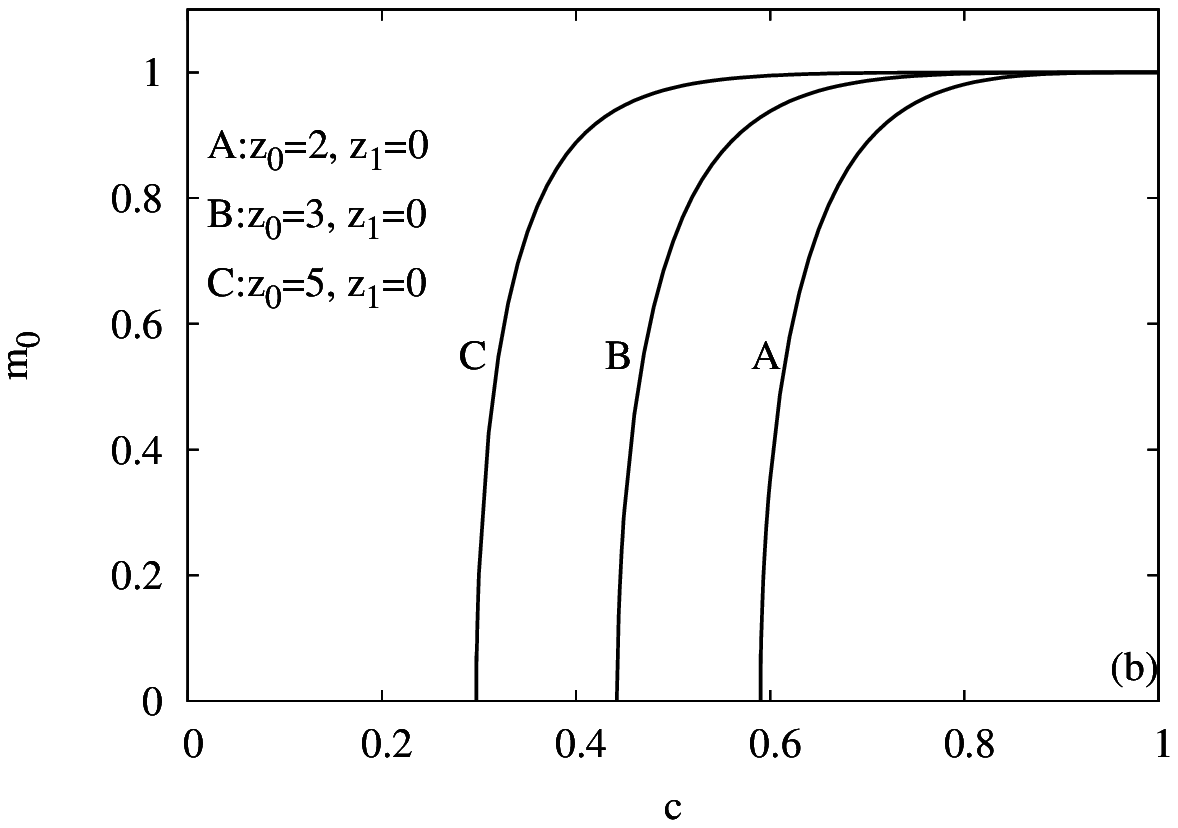, width=7.0cm}
\end{center}
\caption{(a) The variation of the critical temperature
($k_BT_c/J$) of the system with the bond concentration ($c$), (b) the variation of the ground state magnetization of the system with the bond concentration ($c$),
 for selected lattices with the isotropic quantum Heisenberg model.
The fixed parameter values are $r_x=r_y=r_z=1.0$. In (b), the temperature is selected as $k_BT_c/J=0.001$.} \label{sek1}\end{figure}

It is a known fact that for a certain set of Hamiltonian parameters the critical temperature of the pure system with Ising type symmetry is greater than the system with XY type symmetry and this value is greater than that of the system with Heisenberg type symmetry\cite{ref15}.  From this fact we expect that, the relation between the bond percolation threshold values of three types of symmetries to be $c^{*}(H)>c^{*}(XY)>c^{*}(I)$. Here, the letters in the parenthesis depict the symmetry type of the Hamiltonian. Namely, $H$  stands for Heisenberg type symmetry (with the anisotropy values $r_x=r_y=r_z=1.0$), $XY$  stands for XY type symmetry (with the anisotropy values $r_x=0.0, r_y=r_z=1.0$),  and $I$  stands for Ising type symmetry (with the anisotropy values  $r_x=r_y=0.0, r_z=1.0$). But the formulation gives this relation as $c^{*}(XY)>c^{*}(H)>c^{*}(I)$, as seen in Table (1). We will touch upon this point in the next subsection.
The  bond percolation threshold values of the different lattices for different Hamiltonian symmetries can be seen in Table (1).

\begin{table}[h]\label{table1}
\begin{center}
\begin{threeparttable}
\caption{The bond percolation threshold values for several lattices with the Hamiltonians that have Ising type symmetry (I), XY type symmetry (XY) and
Heisenberg type symmetry (H).}
\renewcommand{\arraystretch}{1.3}
\begin{tabular}{lllllllll}
\thickhline
Lattice & $z_0$& $z_1$ &$c^{*}$ (I)&$c^{*}$ (XY)&$c^{*}$ (H)\\
\hline
Honeycomb & 2& 0 & 0.5707&0.6448&0.5903 \\
Kagome & 2& 1 &  0.4655&0.5092&0.4709 \\
Square& 3& 0 & 0.4295 &0.4699&0.4427\\
Triangular& 3& 2 & 0.3102 & 0.3289& 0.3127\\
Simple cubic& 5& 0 & 0.2903 & 0.3085 &0.2974\\
Body centered cubic& 7& 0 & 0.2200 & 0.2305 &0.2245\\
Face centered cubic& 7& 4 & 0.1537  &0.1584 &0.1546\\
\thickhline \\
\end{tabular}
\end{threeparttable}
\end{center}
\end{table}

The effect of the bond concentration on the behavior of the magnetization with temperature can be seen in Figs. \re{sek2}(a) for a honeycomb lattice and \re{sek2}(b) for a simple cubic lattice. As seen in Fig.\re{sek2} that, decreasing $c$ causes a decline in the critical temperature value, also after some value of it, it causes decline in  the ground state magnetization. Ground state magnetization starts to decrease at a lower value of $c$ for the simple cubic lattice (in comparison with the honeycomb lattice), as seen in Fig. \re{sek2} in compatible with Fig. \re{sek1}(b) (e.g. all curves of $c$ value that provide  $0.6 \le c\le 1.0$ for the simple cubic lattice have ground state magnetization $m_0=1.0$ while for the honeycomb lattice the same ground state magnetization value has been observed only for $c=1.0, 0.9$ values). Again, this is due to the excess number of nearest neighbor spins of the simple cubic lattice in comparison with those of the honeycomb lattice.

\begin{figure}[h]\begin{center}
\epsfig{file=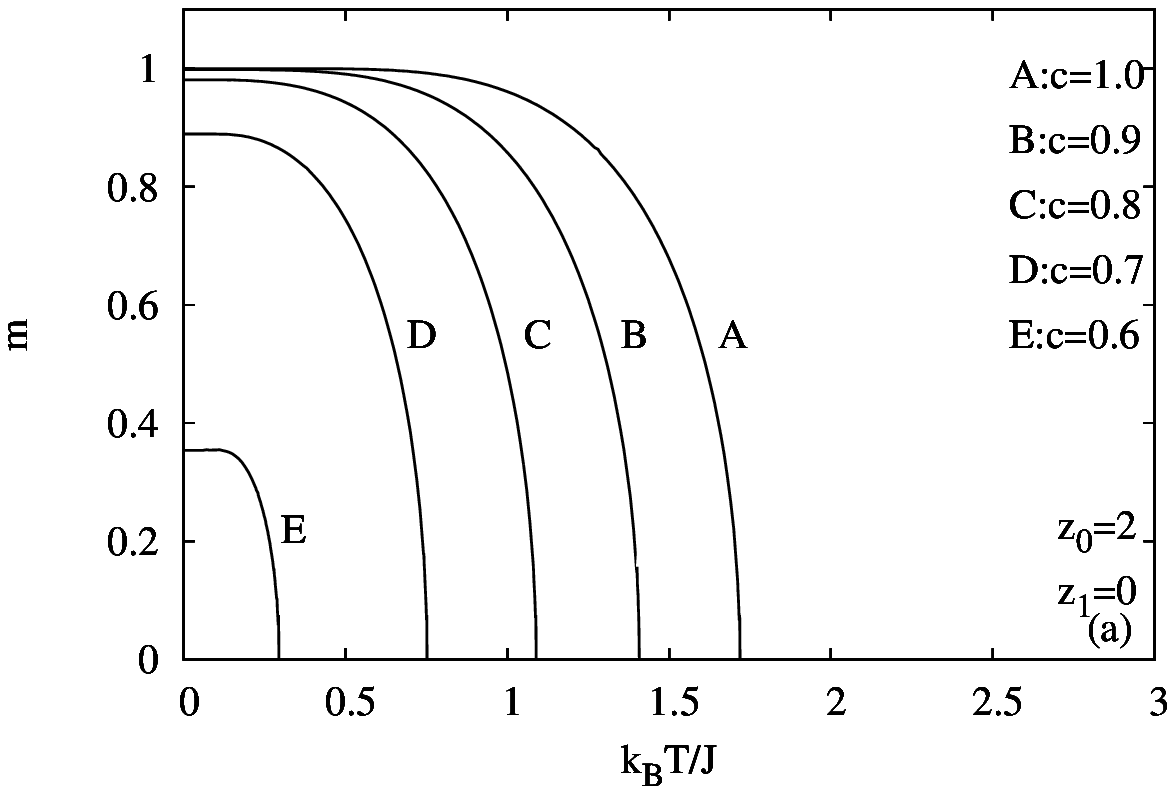, width=7.0cm}
\epsfig{file=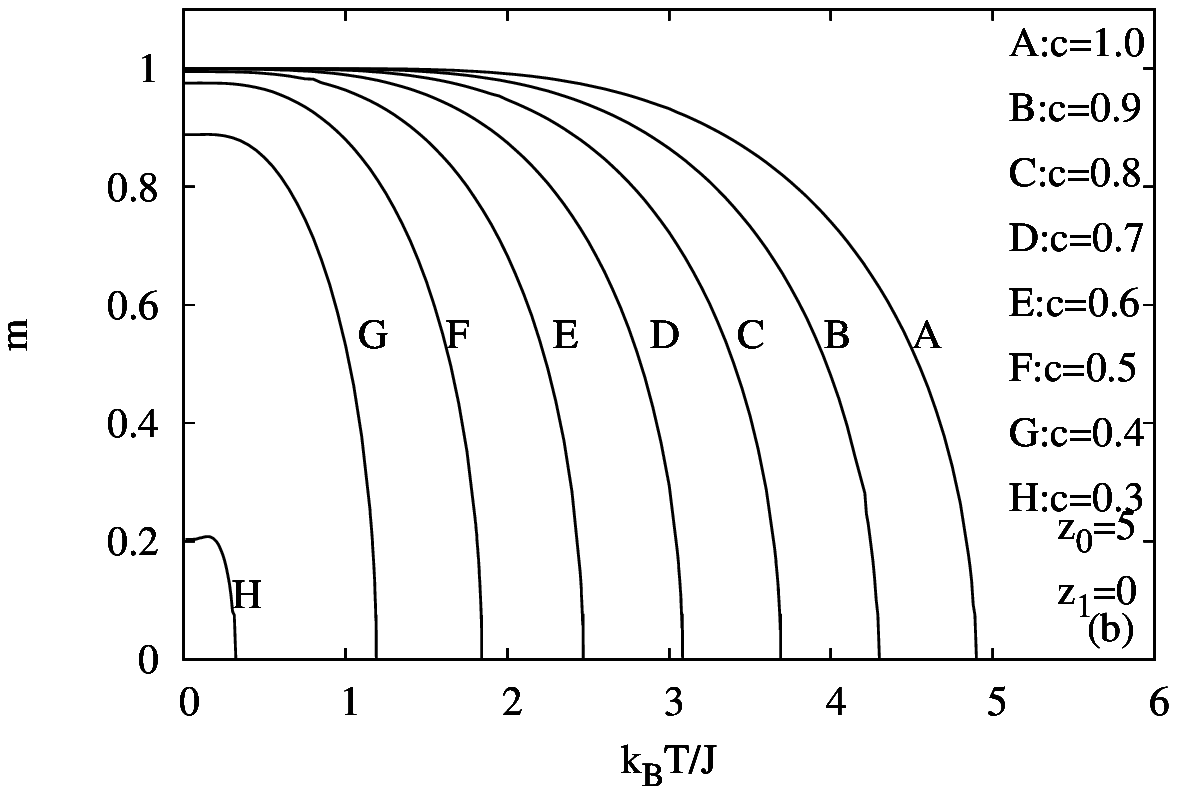, width=7.0cm}
\end{center}
\caption{Variation of magnetization with temperature for (a) honeycomb and (b) simple cubic lattices for selected bond concentration values within the quantum isotropic Heisenberg model. The fixed parameter values are $r_x=r_y=r_z=1.0$.} \label{sek2}\end{figure}

\subsection{Anisotropic Model}

\begin{figure}[h]\begin{center}
\epsfig{file=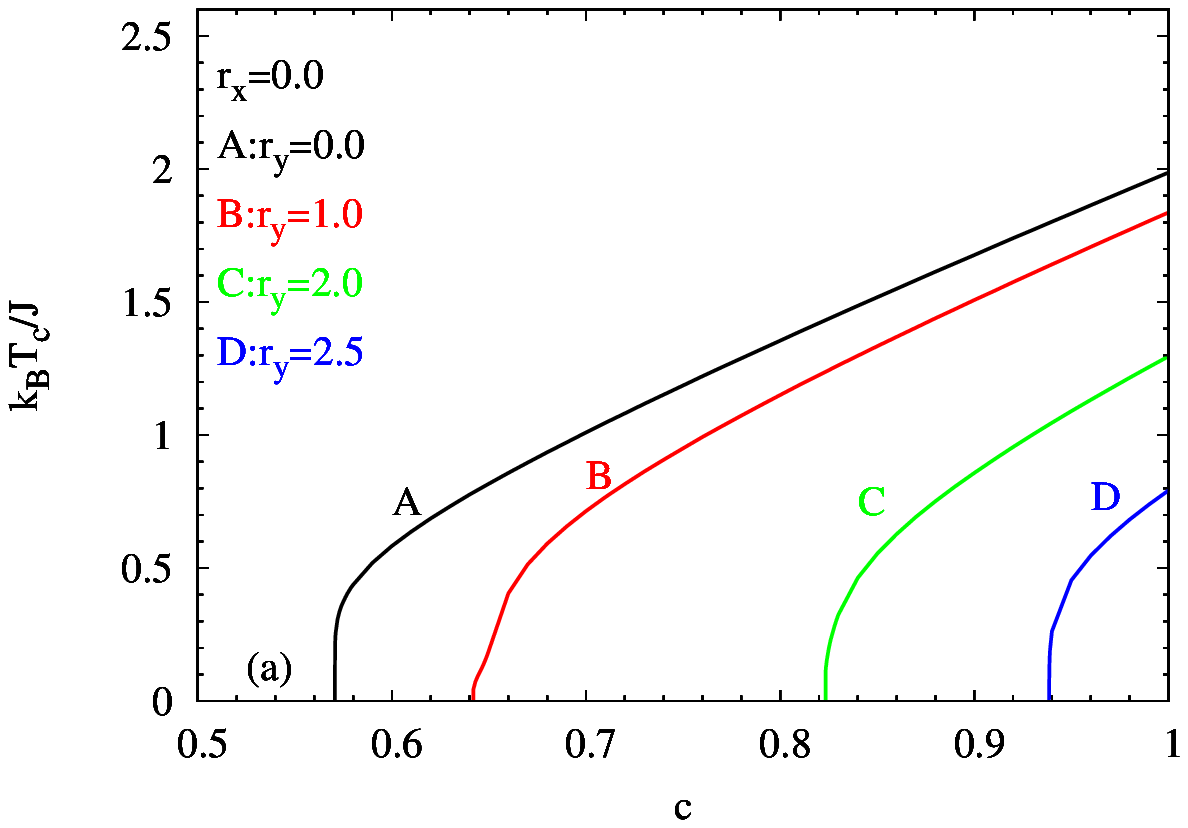, width=7.0cm}
\epsfig{file=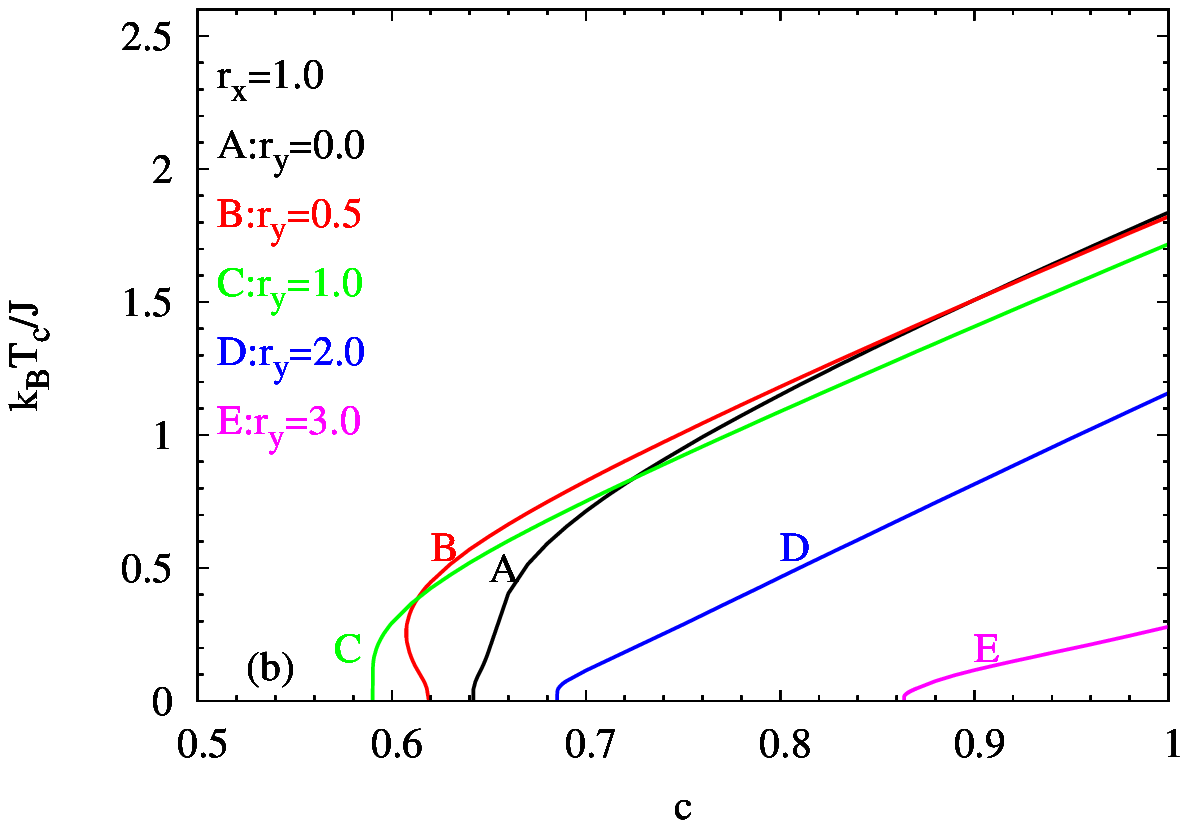, width=7.0cm}

\epsfig{file=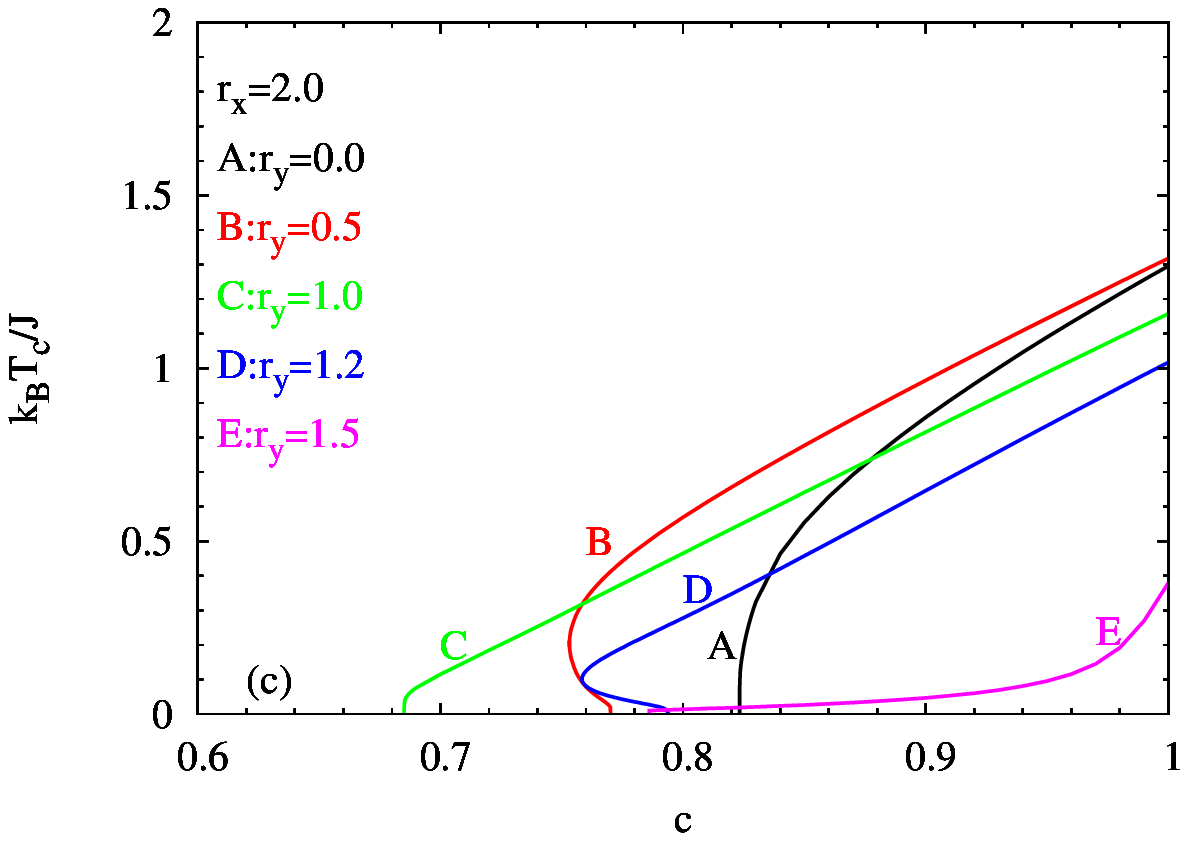, width=7.0cm}
\epsfig{file=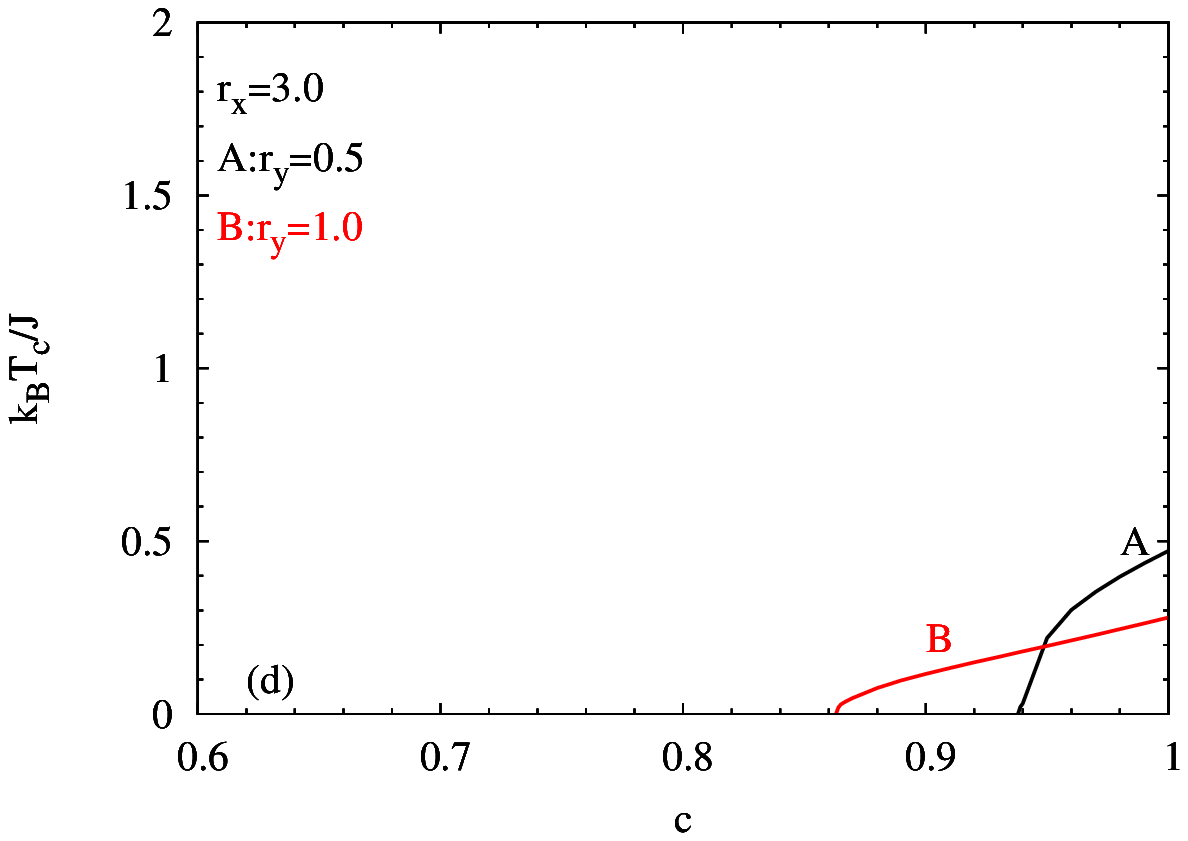, width=7.0cm}
\end{center}
\caption{Variation of the critical temperature ($k_BT_c/J$) of the anisotropic quantum Heisenberg model as a function of bond concentration (c) for a honeycomb lattice, with some selected values of anisotropy in the exchange interaction ($r_x,r_y$). The fixed parameter value is $r_z=1.0$ }  \label{sek3}\end{figure}

Now, let us see the effect of the anisotropy in the exchange interaction on the results obtained for the isotropic case in the
preceding subsection. To do this, we start with the phase diagrams corresponding Fig. \re{sek1} (a)  for the isotropic case. The phase diagrams for a honeycomb lattice in ($k_BT_c/J,c$) plane can be seen in Fig. \re{sek3} for different anisotropy values. In Fig. \re{sek3}(a), we can see the phase diagrams of the system with Ising type symmetry (the curve labeled by A) and XY type symmetry (curves labeled by B,C,D). As we can see from the curves labeled by B,C,D in Fig. \re{sek3}(a), as the amount of anisotropy increases then the ferromagnetic region gets narrower in the ($k_BT_c/J,c$) plane for the XY type model and after a certain value of $c$, phase diagrams disappears. This value depends on the lattice geometry. Bond percolation threshold values of the XY type system gradually decrease when the anisotropy increases. There is not any reentrant behavior observed in the phase diagrams of the XY type system. In Figs. \re{sek3} (b) and (c), transition from the curves labeled by A to the other curves corresponds to the transition from the system with XY type symmetry to the system with H type symmetry. When we take a look at Fig. \re{sek3} (b), we see that increasing anisotropy effects the phase diagrams in the ($k_BT_c/J,c$) plane in a more complicated way in comparison with Fig. \re{sek3} (a). Firstly, as the anisotropy increases then the bond percolation thresholds decrease for a while then tend to increase. We can clearly see here, although the critical temperature of the pure system for the XY type symmetry (the curve labeled A in Fig. \re{sek3}(b)) is greater than that of the H type symmetry (curve labeled C in Fig. \re{sek3}(b)), this does not imply that the bond percolation threshold value of the XY type symmetric model to be lower than the H type symmetric model. This behavior of the curves labeled A and C seen in Fig. \re{sek3}(b) verifies the results in Table (1). Secondly, increasing anisotropy can induce a reentrant behavior in the phase diagrams (e.g. see the curve labeled by B in Fig. \re{sek3} (b)). Besides, we can see from Fig. \re{sek3} (c) that, increasing anisotropy in the exchange interaction can produce phase diagrams that behave qualitatively different from the others. For instance, for the curve labeled by E in Fig. \re{sek3} (c), decreasing anisotropy first causes a sharp decrease in the critical temperature then the critical temperature stays almost constant before it depresses to zero. In this figure, the curves labeled by B and D also exhibit reentrant behavior. One other interesting behavior can be seen in Fig. \re{sek3} (d). For $r_x=3.0$, with low anisotropies (i.e. small values of $r_y$), there are no phase diagrams. The system can not exhibit an ordered phase. When the value of $r_y$ increases then the phase diagrams appear and then disappear again. In conclusion, we can say that while the transition from the $I$ type symmetry to the $XY$ type symmetry (Fig. \re{sek3} (a)) rising anisotropy in the exchange interaction can not induce reentrant behavior in the phase diagrams which are in the $(k_BT_c/J,c)$ plane. But, if increasing anisotropy causes a transition from the $XY$ type symmetry to the $H$ type symmetry then it can induce a reentrant behavior in the phase diagrams in the $(k_BT_c/J,c)$ plane, (Fig. \re{sek3} (b) and (c)).

In order to concentrate on the evolution of the bond percolation threshold value with anisotropy in the exchange interaction, let us investigate the variation of $c^*$ value with $r_y$ for some selected values of $r_x$. These curves can be seen in Fig. \re{sek4}(a). The curve labeled by A corresponds to the XY type system. As mentioned in the comments of Fig. \re{sek3}, for the XY type system, with increasing anisotropy, the bond percolation threshold value smoothly increases to the value $1.0$ while for the system with H symmetry, the bond percolation threshold value first decreases with increasing anisotropy then increases and reaches to the value of $1.0$ (see the curves labeled by B,C,D in Fig. \re{sek4}(a)). When $r_x$ increases, the $c^*$ value corresponding to $r_y=0$ increases and after a specific value of $r_x=2.7105$ the $c^*$ value for $r_y=0$ becomes $1.0$. This value is nothing but the value of the curve labeled by A intersects the $c^*=1.0$, due to the symmetry of the formulation as mentioned at the beginning  of this section. This is just the value that makes the critical temperature of the pure system (i.e. $c=1.0$) equal to zero with $XY$ type symmetry. We can numerically determine the similar borders of the model with $H$ type symmetry.

As seen in  Fig. \re{sek4} (a), curves intersect the $c^*=1.0$ line at a specific value of $r_y$ which depends on the value of $r_x$. The curves labeled A,B and C intersect the $c^*=1.0$ line only by one point while, the curve labeled D by two point. Let denote these two intersection points as $r_l^*$ and $r_r^*$ which provide  $r_r^*>r_l^*$. In other words  $r_r^*$ denotes the value of $r_y$ which is related $r_x$ curve intersects to the $c^*=1$ line on the right side and $r_l^*$ denotes the value of $r_y$ which related $r_x$ curve intersects to the $c^*=1$ line on the left side, if present. These two special values in the $(r_x,r_y)$ plane can be seen in Fig. \re{sek4} (b). For any point which is outside of the region confined by this curve and $r_x,r_y$ axes, we can say that the system can not exhibit an ordered phase for a honeycomb lattice. This border consists of three segments: $r_y=r_x+2.7105$ line, $r_y=r_x-2.7105$ line and the remaining curve. The region that the system can exhibit an ordered phase becomes expanded when $r_x$ increases from $0.0$ to $1.0$, after then that region starts to shrink. We can say from the symmetry properties of the formulation that, this border curve is symmetric about the $r_y=r_x$ line. We can obtain similar borders for the system which has bond concentration values different from $c= 1.0$. These curves can be seen in   Fig. \re{sek4} (c). Again we can say with Fig. \re{sek4} (c) that, for any point which is outside of the region confined by the related $c$ curve and $r_x,r_y$ axes, the system with $(1-c)$ percentage open bonds, can not exhibit an ordered phase for a honeycomb lattice.

Finally, we want to investigate the effect of the bond dilution on the ground state magnetizations of the system. In the isotropic case, this effect was as decreasing $c$ gradually decreases the ground state magnetizations (Fig. \re{sek1} (b)). But in anisotropic case, this situation is a little complicated as seen in Fig. \re{sek5}, which is the  variation of the ground state magnetization ($m_0$) with anisotropy ($r_y$) and bond concentration ($c$) for some selected values of $r_x$. Fig. \re{sek5} (a) is just concerning the model with $XY$ type symmetry. There is not any complicated behavior as in the isotropic model. For a certain $c$, the ground state magnetization regularly decreases with increasing $r_y$.

But the situation is more complicated in Figs. \re{sek5} (b) and (c). These are the surfaces regarding the model with $H$ type symmetry and all the reentrant behaviors in the phase diagrams of the system (e.g. curve labeled by $B$ in Fig. \re{sek3} (b) and  (c)) shows itself in the behavior of the ground state magnetizations. Beside these, $r_x=2.0$ surface is not smooth while $r_x=1.0$ surface is smooth i.e., for $r_x=1.0$ with any value of $c$, increasing $r_y$ changes the ground state magnetization continuously but this is not the case for $r_x=2.0$. Increasing $r_x$ value also shrinks this surface as expected. Fig. \re{sek5}(d) has different characteristics than the others; the ground state magnetization surface can not reach the small valued $r_y$ regions, as well as the higher valued $r_y$ regions. This fact is compatible with the curve labeled by D in Fig. \re{sek4} (a).

\begin{figure}[h]\begin{center}
\epsfig{file=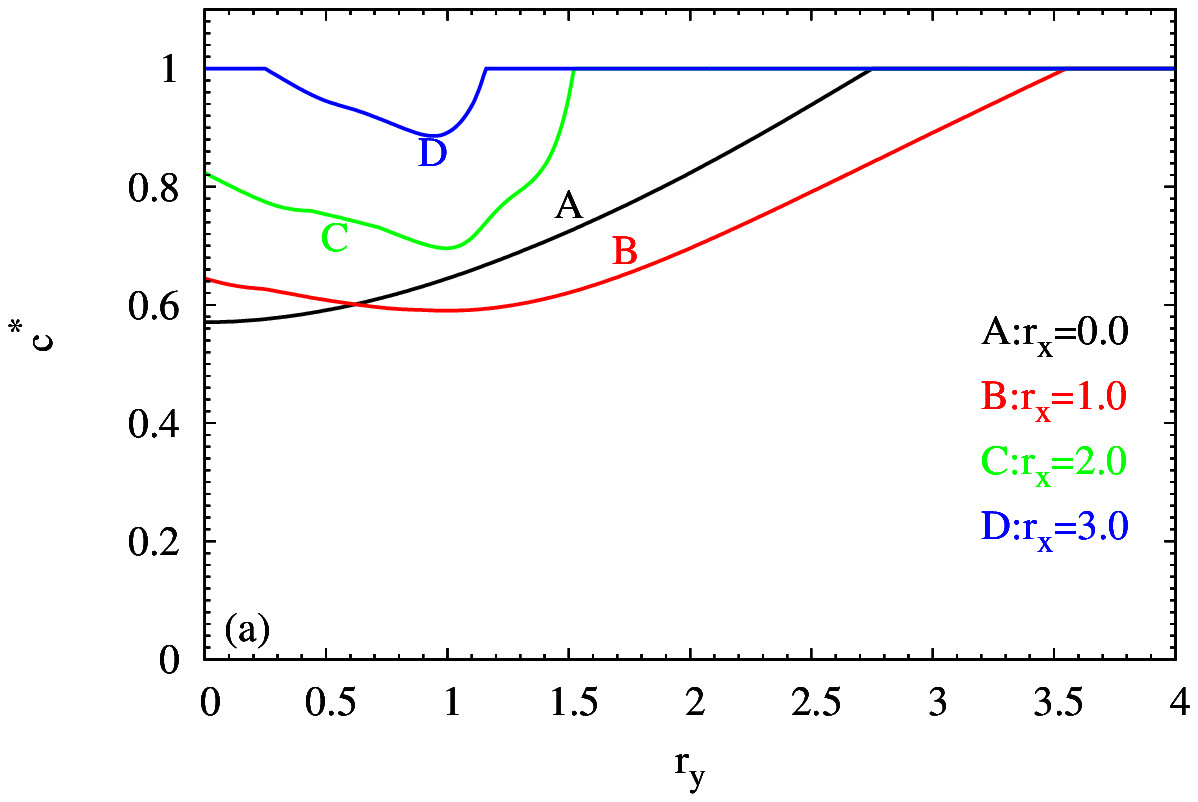, width=7.0cm}
\epsfig{file=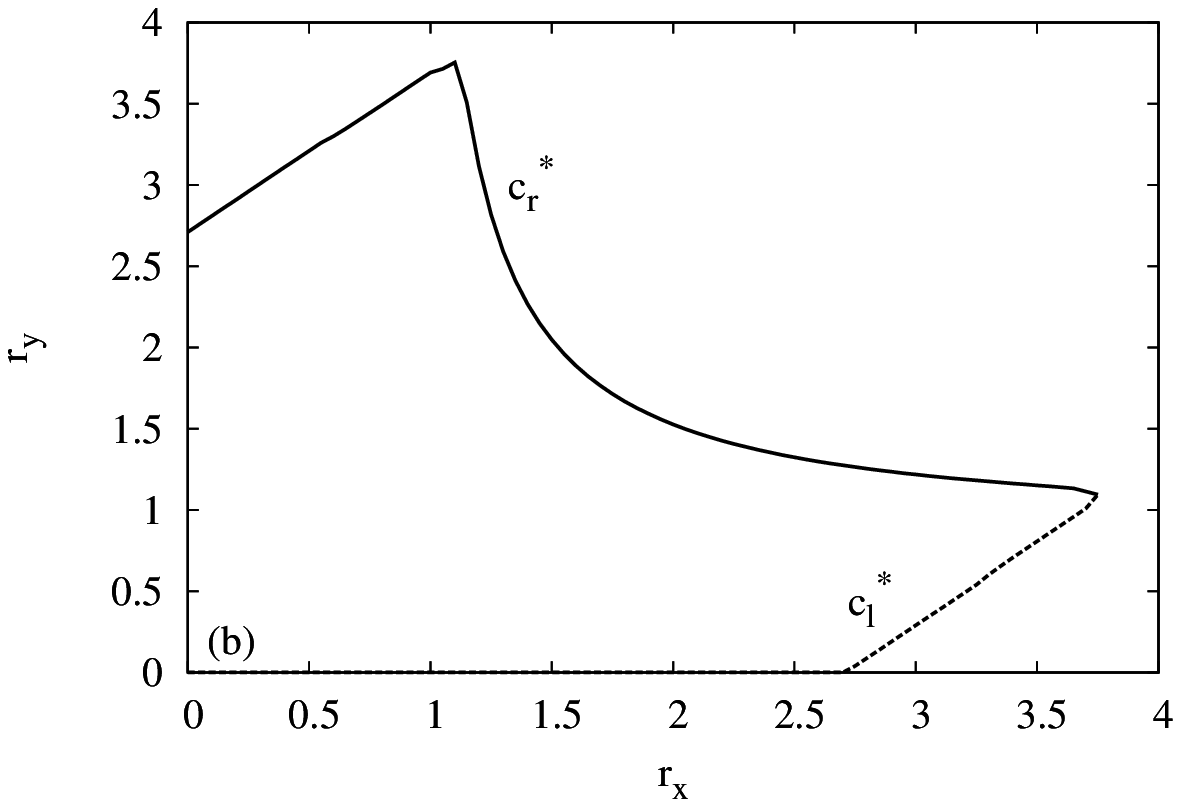, width=7.0cm}
\epsfig{file=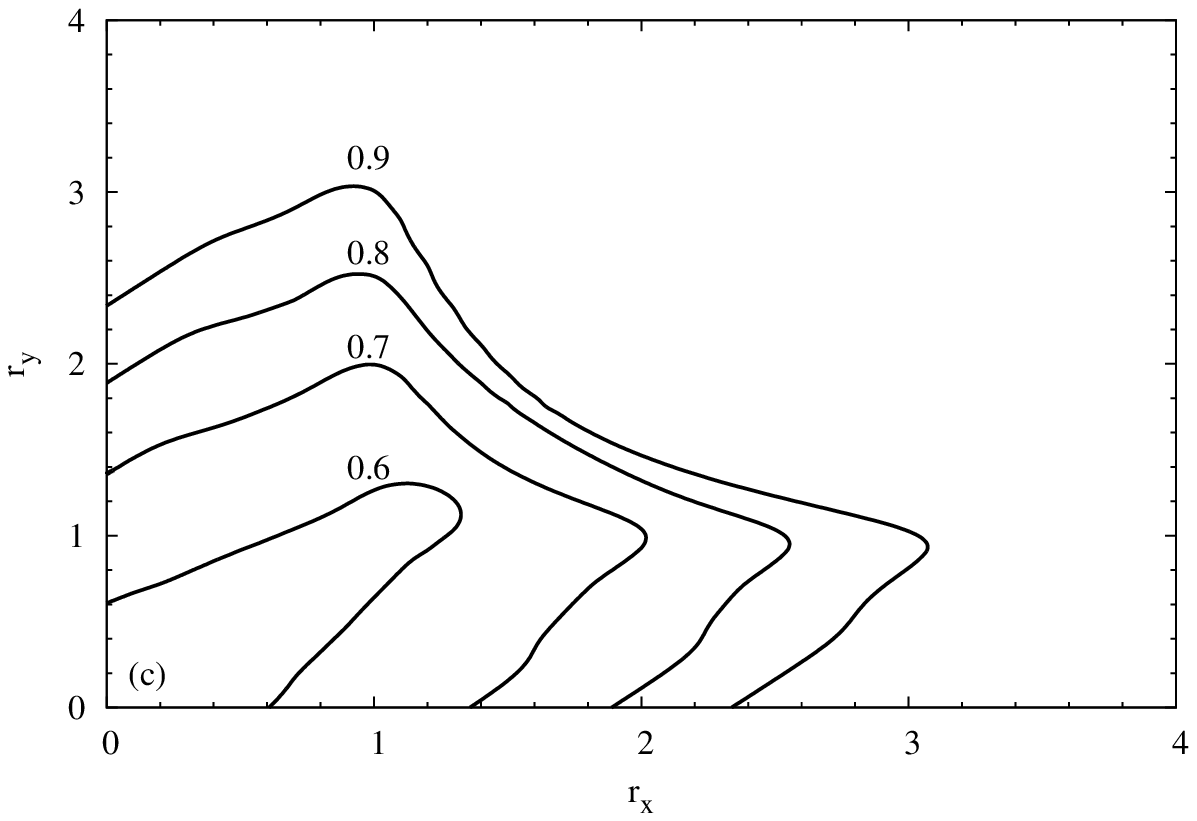, width=7.0cm}
\end{center}
\caption{(a) Variation of the bond percolation threshold value ($c^*$) with $r_y$ for some selected values of $r_x$,
(b) the border of the ferromagnetic region of the pure ($c=1.0$) model in the $(r_x,r_y)$ plane,
(c) equally valued  bond percolation threshold curves in the $(r_x,r_y)$ plane for a honeycomb lattice within anisotropic quantum Heisenberg model.
The fixed parameter value is $r_z=1.0$.} \label{sek4}\end{figure}

\begin{figure}[h]\begin{center}
\epsfig{file=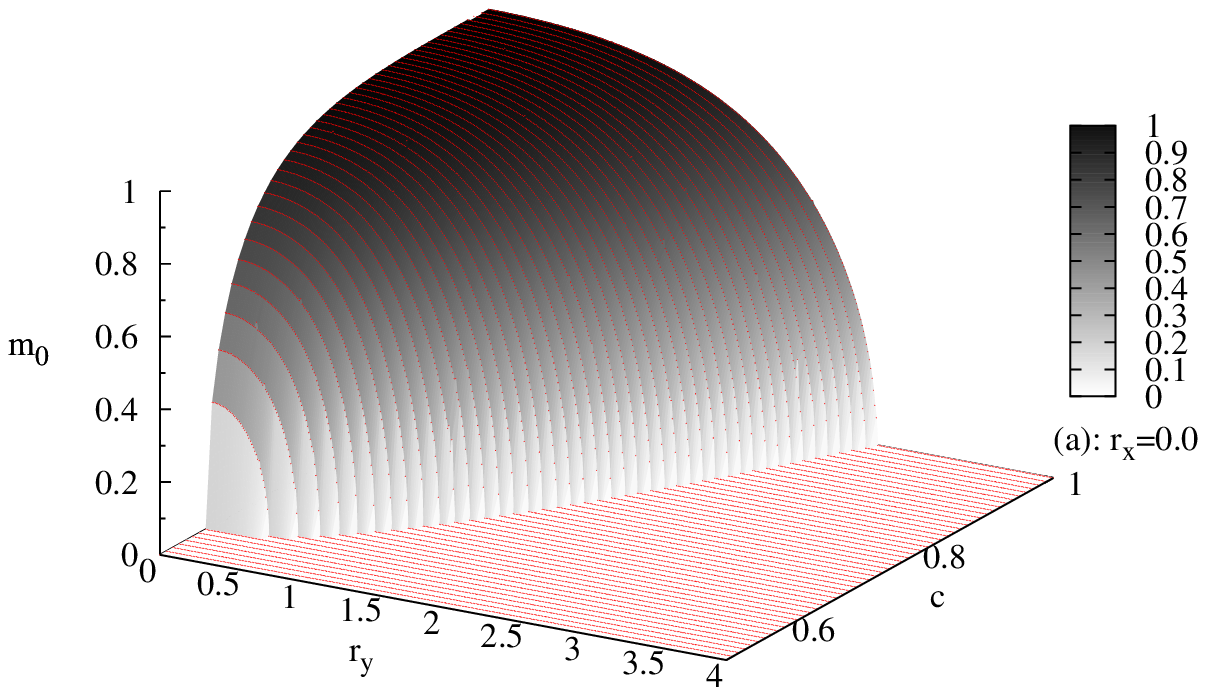, width=7.0cm}
\epsfig{file=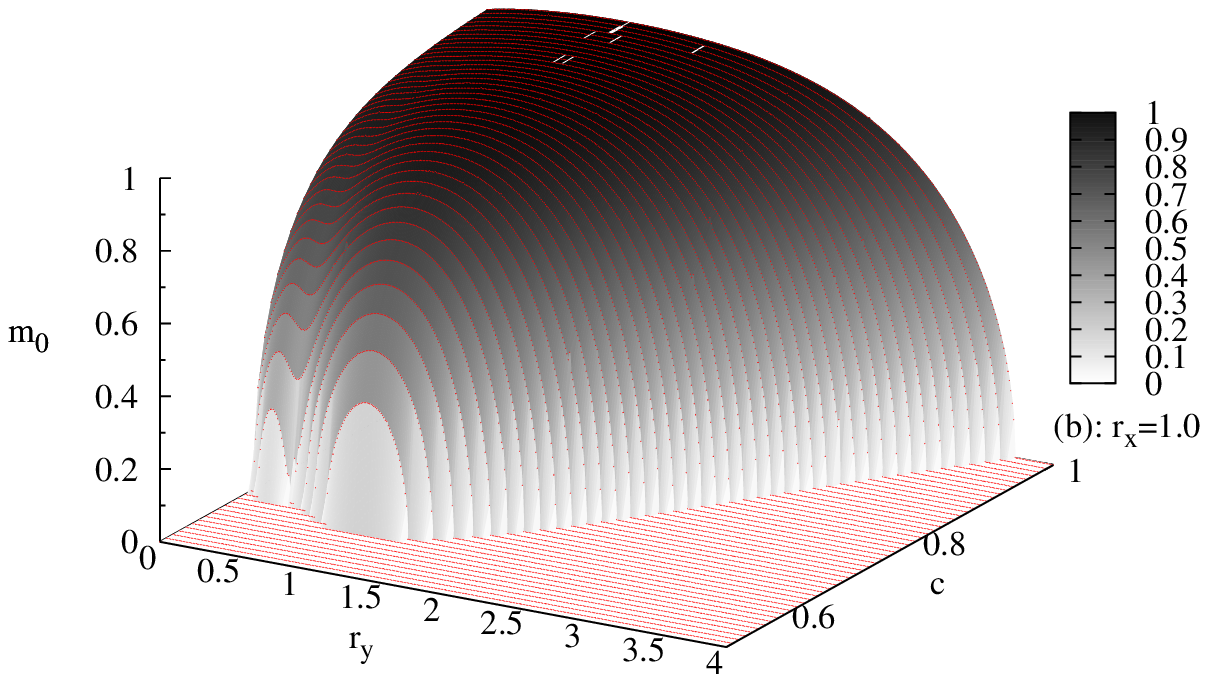, width=7.0cm}

\epsfig{file=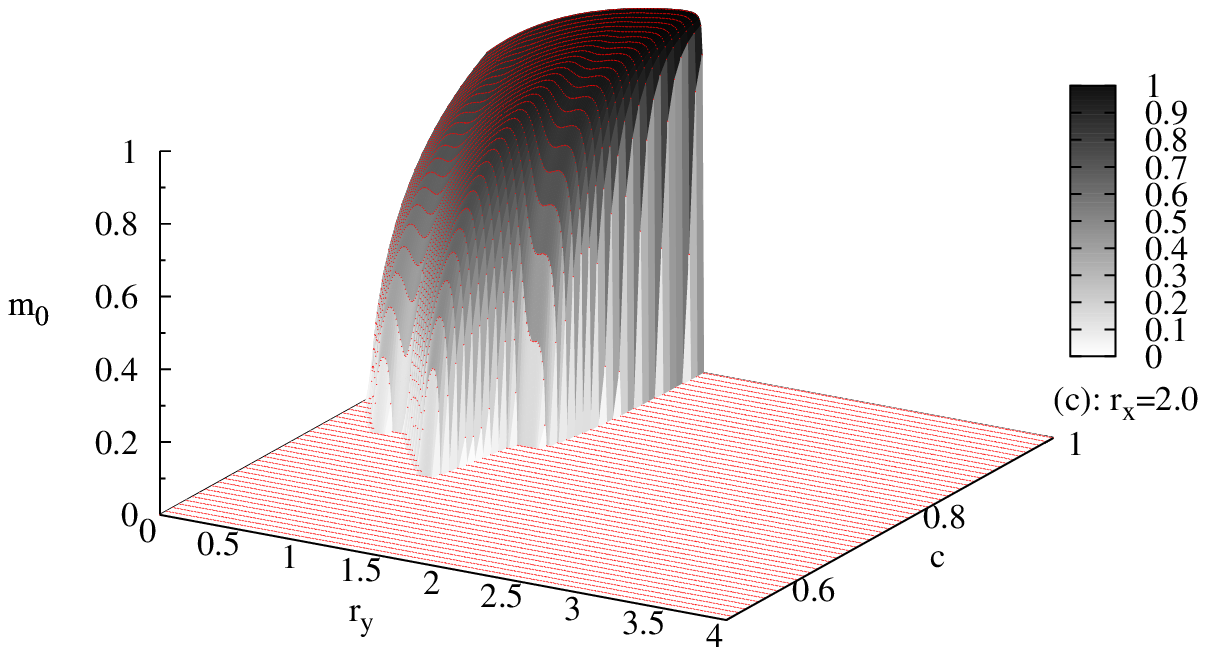, width=7.0cm}
\epsfig{file=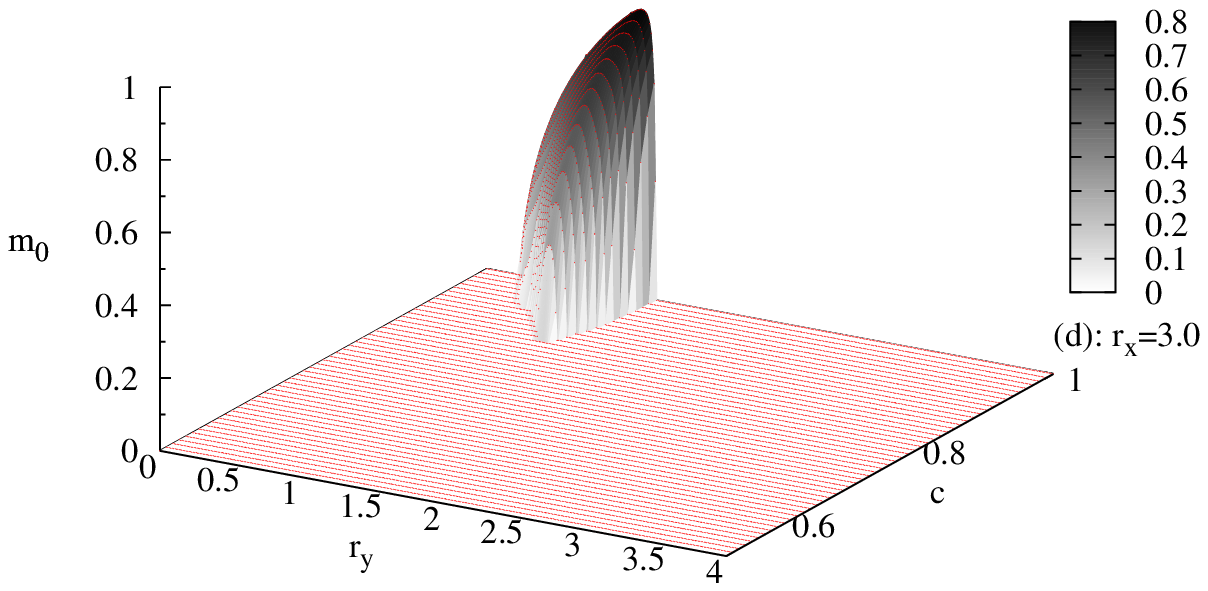, width=7.0cm}
\end{center}
\caption{Variation of the ground state magnetization with $c$ and $r_y$ for a honeycomb lattice within the quantum Heisenberg model for some selected values of $r_x$.
The fixed parameter value is $r_x=1.0$.} \label{sek5}\end{figure}

\section{Conclusion}\label{conclusion}

The effect of the bond dilution on the phase
diagrams of the isotropic and anisotropic quantum Heisenberg model has been
investigated in detail. First in the isotropic model, effect of the bond dilution process on the critical temperatures have been given for honeycomb, square and simple cubic lattices. There is not any reentrant behavior observed in that phase diagrams plotted in ($k_BT_c/J,c$) plane for the isotropic Heisenberg model. Also, the variation of the ground state magnetization with bond concentration ($c$) and the evolution of the variation of the magnetization with temperature curves with $c$ have been presented. The bond percolation threshold values which is the value of $c$ that terminates the phase diagrams in the $(k_BT_c/J,c)$ plane have been given for several 2D and 3D lattices with the Hamiltonians that have different symmetries.

In addition,  the same investigation has been performed for the anisotropic model. Anisotropic model has been treated within three different regions. Namely, the Hamiltonian that have Ising type symmetry $r_x=r_y=0.0, r_z=1.0$, XY type symmetry $r_x=0.0, r_y>0.0, r_z=1.0$ and Heisenberg type symmetry $r_x, r_y>0.0, r_z=1.0$. All these models that have different symmetry properties have been investigated in detail: The effect of the bond dilution process on both the ground state magnetizations and critical temperatures have been determined via the phase diagrams in the  $(k_BT_c/J,c)$ plane. Beside these, the bond percolation threshold values ($c^*$) have been determined for the anisotropic model and depicted in the $(c^*,r_y)$ plane for selected values of $r_x$. For the XY type symmetric model, decreasing bond concentration results in gradually increase of the bond percolation threshold value. There is not any reentrant behavior observed within the model that have XY type symmetry on the phase diagrams in the   $(k_BT_c/J,c)$ plane.

On the other hand, in the model that have Heisenberg type symmetry, this situation is more complicated. Increasing anisotropy does not results in a gradually increase of the $c^*$ value as in the model that have XY type symmetry. However, increasing anisotropy first causes a decrease in the $c^*$ value then an increase in it before it reaches the value of $1.0$. Beside this fact, increasing anisotropy also give rise to the reentrant behavior in the phase diagrams in the $(k_BT_c/J,c)$ plane. All of these reentrant behaviors are found to be of the second order.

We hope that the results  obtained in this work may be beneficial form both theoretical and experimental point of view.

\end{document}